\documentclass[a4paper,12pt]{article}

\usepackage[authoryear]{natbib}
\usepackage{authblk}
\usepackage{amsfonts,amstext,amssymb,amsmath}
\usepackage{calrsfs}
\usepackage[svgnames,dvipsnames]{xcolor}
\usepackage{multicol}
\usepackage{dblfloatfix}
\usepackage{subfig}
\usepackage{graphicx}
\usepackage{nicefrac}
\usepackage{physics}
\usepackage{todonotes}
\usepackage{ulem}

\usepackage{tabularx}
\usepackage{booktabs}
  
\usepackage{hyperref}
\usepackage{wrapfig}

\usepackage[margin=2cm]{geometry}

\usepackage[capitalise,nameinlink]{cleveref}

\hypersetup{colorlinks={true},linkcolor={blue},citecolor=blue}

\DeclareMathAlphabet{\pazocal}{OMS}{zplm}{m}{n}

\newcommand{\addition}[1]{\textcolor{black}{#1}}

\title{Wrinkling instability of 3D auxetic bilayers in tension}                      

\author[1]{Sairam Pamulaparthi Venkata}
\affil[1]{School of Mathematical and Statistical Sciences, University of Galway, Galway, Ireland}

\author[2]{Yuxin Fu}
\affil[2]{Department of Mechanics, Tianjin University, Tianjin, People’s Republic of China}

\author[3]{Yibin Fu}
\affil[3]{School of Computer Science and Mathematics, Keele University, United Kingdom}

\author[4]{Hooman Danesh}
\affil[4]{Institute of Applied Mechanics, RWTH Aachen, Aachen, Germany}

\author[1,5]{Michel Destrade}
\affil[5]{Department of Engineering Mechanics, Zhejiang University, Hangzhou, People’s Republic of China}

\author[1]{Valentina Balbi}

\date{}                     
\begin{document}
  \maketitle

\begin{abstract}
Bilayers, soft substrates coated with stiff films, are commonly found in nature with examples including skin tissue, vesicles, and organ membranes. 
They exhibit different types of instabilities when subjected to compression, depending on the contrast in material properties between the two components.
In this work, we unravel the mechanisms behind wrinkling instabilities in auxetic bilayer systems under uniaxial tension.
We find that a soft bilayer in tension can experience significant lateral contraction, and with sufficient contrast in Poisson ratios, compressive stresses may induce wrinkles aligned with the tensile direction.
We analytically model the onset of wrinkles and validate our predictions using Finite Element simulations in \textsf{ABAQUS}.  
Our findings reveal that wrinkles occur when the Poisson ratio of the substrate is greater than that of the film. As the two Poisson ratios converge to a common value, the critical stretch for instability shoots up rapidly and the wrinkles disappear. 
We also confirm these results through asymptotic analysis. Using inverse analysis, we design film microstructures to achieve desired effective Poisson ratios and further validate the effective properties with the Finite Element code \textsf{FEAP}. We show that the critical stretch ratio for buckling in auxetic structures with microstructural patterns is in strong agreement with the homogenized model predictions. The proposed method has significant potential for controlling surface patterns in auxetic skin grafts and hydrogel organ patches under mechanical loads. Moreover, the asymptotic expressions for compressible bilayers developed in this work can also be applied under finite strain for buckling-based metrology. 

\end{abstract}


\section{Introduction}\label{intro}

\begin{figure*}[!th]
\centering
\includegraphics[width=0.7\textwidth]{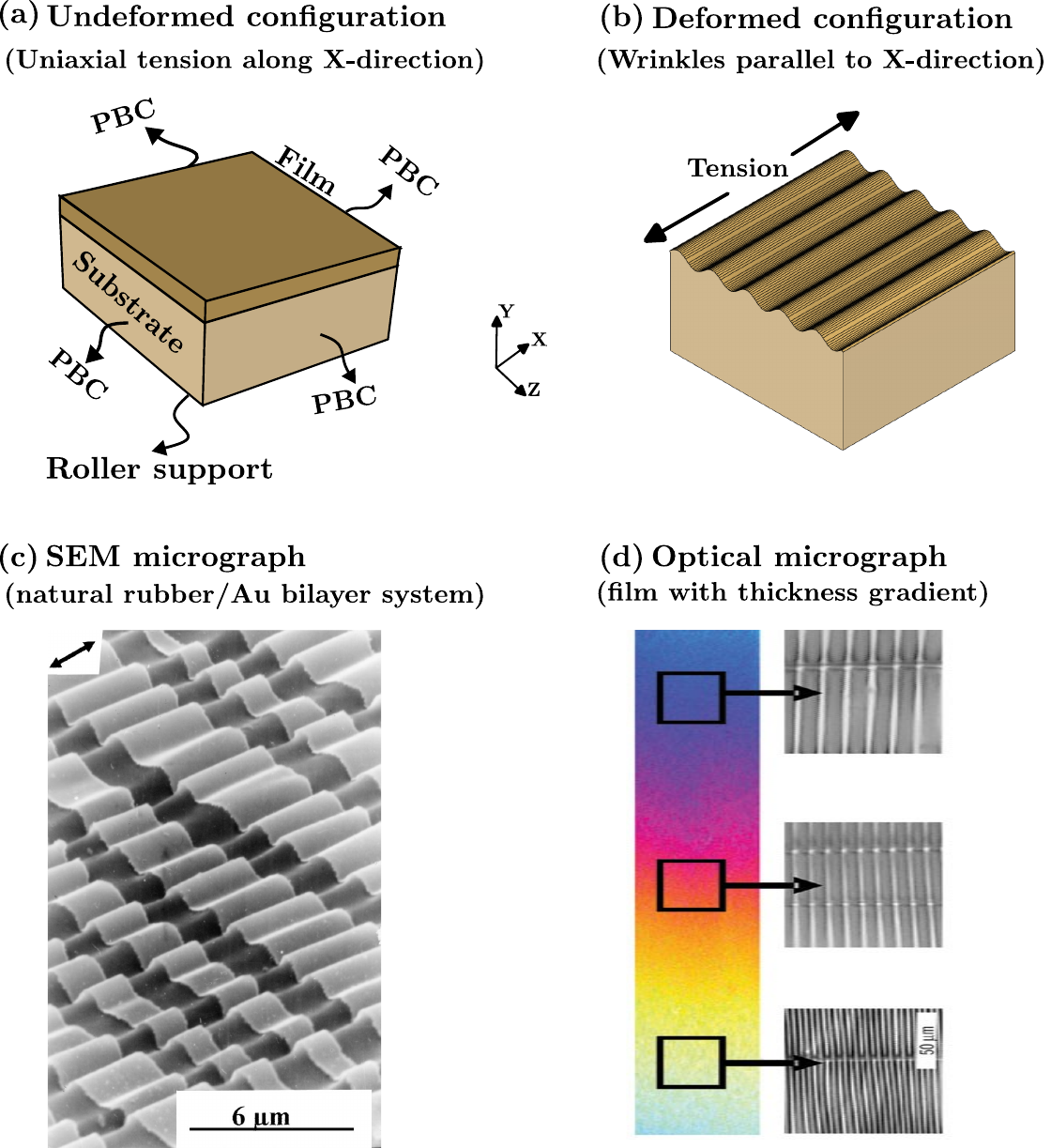}
    \caption{(a) Undeformed and (b) deformed configurations of a 3D bilayer system with periodic boundary conditions (PBCs) in the lateral directions. The system is infinite in the X- and Z-directions and is under uni-axial tension along the X-direction. The analysis shows that eventually, (b) wrinkles develop parallel to the X-direction, provided there is enough contrast between the constitutive parameters of the materials. (c) Image obtained using Scanning Electron Microscope (SEM) on a natural rubber substrate coated with a gold film, when elongated to 50\% strain at room temperature \citep{Volynskii2000}. An inset with an arrow shows the direction of elongation. Bright and dark bands represent the buckled gold film layer and the rubber showing through cracks in the stretched film, respectively. (d) Optical micrograph of a Polystyrene (PS) film layer on a silicon wafer with decreasing thickness from top to bottom is shown on the left panel. The corresponding wrinkles exhibit decreasing wavelengths on the right panel, when the PS films are attached to a Polydimethylsiloxane (PDMS) substrate and elongated to induced buckling \citep{Stafford2004}.}
    \label{fig:3dmodel}
\end{figure*}

The formation of wrinkles on the surface of a solid under applied tension is rarely studied beyond the linear elastic framework \citep{Siavash20192}, although they have been observed experimentally under large strains \citep{Volynskii2000,Stafford2004}, see \cref{fig:3dmodel} of illustrations and examples.
In this paper, we conduct a linear buckling analysis of 3D hyperelastic thin stiff films on semi-infinite compliant substrates, including auxetics, under tension and large strains, addressing a gap in the literature, which has, so far, primarily focused on compression. 
We refer the reader to \cref{AppendixA} for short literature recaps on the wrinkling instability of layered media and on auxetic materials.

To conduct this investigation, we develop a semi-analytical approach in \textsf{Mathematica} \citep{Mathematica} to predict the onset of wrinkles in bilayers under uniaxial tension. 
We also create a UHYPER subroutine in FORTRAN to implement appropriate constitutive models in \textsf{ABAQUS} \citep{ABAQUS}, and use custom Python scripts to simulate wrinkling in compressible bilayers with periodic boundary conditions using Finite element (FE) analysis. 

Key novelties include 
(i) Deriving asymptotic expressions for the critical stretch ratios and wavenumber of wrinkling, applicable under finite strains and relevant for buckling-based metrology; 
(ii) Demonstrating that wrinkles occur only when the Poisson ratio of the substrate exceeds that of the film (as the Poisson ratios converge, the critical stretch for the instability increases sharply, causing the wrinkles to disappear). 
(iii) Validating the semi-analytical approach and asymptotic expressions with detailed FE simulations; 
(iv) Achieving targeted effective Poisson ratios in the film by altering its microstructure with an inverse design approach, and finding that the critical stretch ratio for auxetic structures with microstructural patterns aligns closely with the predictions of the homogenized models.

The paper is structured as follows. 
\cref{matmodel} presents the theoretical background, including the semi-analytical method, the derivation of asymptotic expressions, and an overview of the homogenization and inverse design analyses. 
\cref{methods} describes the computational approach, where we implement the Blatz–Ko strain energy model in a user-defined subroutine (\textsf{UHYPER}) for \textsf{ABAQUS} and perform buckling analysis of three-dimensional bilayer systems. We also investigate the buckling behavior of bilayers with auxetic film microstructures and compare the results with those obtained from homogenized models. 
Finally, \cref{conclusions} summarizes the key findings and outlines the directions for future research.

\section{Theoretical framework}\label{matmodel}



\subsection{3D bilayer system in tension}
\label{3dmodel1}


We consider 3D bilayer systems with a thin stiff film perfectly bonded to a semi-infinite compliant substrate, see \cref{fig:3dmodel}(a,b).
Under a uniaxial tension applied along the X-direction, the bilayer is intuitively expected to eventually develop wrinkles aligned with that direction, as shown in \cref{fig:3dmodel}(b), provided compressive stresses develop along the Z-direction due to a sufficient Poisson ratio contrast. 

To explore the influence of material properties on this wrinkling behavior, we analyze cases where both the film and substrate exhibit either auxetic or conventional mechanical responses. Auxetic materials, characterized by a negative Poisson ratio, exhibit high compressibility and are distinctly different from nearly incompressible materials with a Poisson ratio close to 0.5.

Earlier studies have employed continuum hyperelastic models derived from the Blatz–Ko strain energy function \citep{blatz1962application} to describe auxetic behavior, as they capture the mechanical response of materials with negative Poisson ratios, although with certain limitations \citep{Crespo2018}.
For example, \citet{ciambella2014continuum} employed the Blatz--Ko model to replicate the experimental observations of \citet{Choi1992} on auxetic materials.
The Blatz--Ko strain energy function is as follows, 
\begin{equation}
W =c_1 \left(I_1-3+\frac{1}{\beta}\left(I_3^{-\beta}-1\right)\right) + c_2 \left(\frac{I_2}{I_3} - 3 + \frac{1}{\beta} \left(I_3^{\beta}-1\right)\right),
\label{eq:energy1}
\end{equation}
where $c_1\!=\!\alpha \frac{\mu}{2}$, $c_2\!=\!(1-\alpha) \frac{\mu}{2}$, $\beta=\frac{\nu}{1-2\nu}$, $\mathbf{F}$ is the deformation gradient, and $I_1=\text{tr}(\mathbf{FF}^T)$, $I_2 = I_3\text{tr}[(\mathbf{FF}^T)^{-1}$], $I_3 = \det(\mathbf{FF}^T)$ are three strain invariants. Also, the material constants are 
the non-dimensional parameter $0<\alpha<1$, the initial shear modulus $\mu>0$, and the Poisson ratio $-1 < \nu \le 1/2$. 

\begin{figure*}[!htb]
\centering
\includegraphics[width=\textwidth]{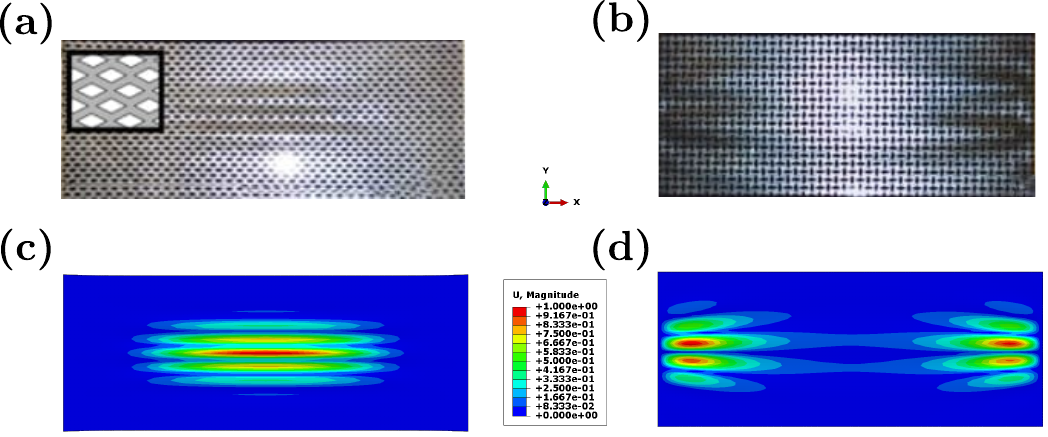}
    \caption{Comparison of wrinkling profiles in thin conventional and auxetic sheets. Top row: Thin acetate sheets under uniaxial tension, with (a) non-auxetic micro-structural patterns and (b) auxetic micro-structural patterns \citep{bonfanti2019elastic}. For conventional sheets, wrinkles develop at the center of the sheet. For auxetic sheets, they appear near the clamped edges.
    Bottom row: Buckling profiles (according to the magnitude of the displacement field) obtained with \textsf{ABAQUS} and the Blatz--Ko model \cref{eq:energy1}. Here $\alpha=0.4$, $\mu = 0.53$ GPa, and $\nu = 0.38, -0.2$ in (c) and (d), respectively.}
    \label{fig:blatzKoexpwrinkl}
\end{figure*}

\cref{fig:blatzKoexpwrinkl} shows that the Blatz--Ko model can effectively capture the experimentally observed behavior of thin membranes (without a substrate) under tension, encompassing both auxetic and conventional Poisson ratios. The model accurately distinguishes the differences in the wrinkling patterns between auxetic and conventional materials, aligning with the experimental findings of \citet{bonfanti2019elastic}, who studied thin acetate sheets with laser-cut microstructural patterns.

To conduct the \textsf{ABAQUS} simulations of \cref{fig:blatzKoexpwrinkl}(c,d), we implemented the Blatz--Ko model using a user-defined hyperelastic subroutine (UHYPER), and then performed a linear buckling analysis to determine the onset of wrinkling. 
Here, instead of explicitly modeling the microstructural patterns, we assigned effective material properties representative of auxetic and conventional behavior to the films. 
This approach allows for a direct comparison between numerical predictions from homogenized models and experimental observations of microstructured models.


\subsection{Semi-analytical treatment: linear buckling analysis}
\label{method1}


To establish a consistent notation for the bilayer system, we denote variables associated with the film and substrate layers using the letters $f$ and $s$, respectively.

To determine the critical stretch and wavelength of the wrinkling instability, we employ the small-on-large method: we first compute the base (elastic) solution, and then superimpose an incremental deformation and derive, and solve, the incremental governing equations and boundary conditions. 

The Blatz--Ko model (\cref{eq:energy1}) is used to obtain the principal components of the Cauchy stress $\mathbf{T}$ as follows:
\begin{equation}
T_{ii} = \frac{E}{2(1+\nu)} J^{-(2\alpha+1)}
\left[-\alpha + [(\alpha-1)\lambda_i^{-2} +\alpha\lambda_{i}^2)]J^{2\alpha} - (\alpha-1)J^{4\alpha}\right], 
\label{eq:cauchystress1}
\end{equation}  
where $\mathbf{F}$ is the deformation gradient, $\lambda_i$ are the principal stretch ratios, and $J = \det \mathbf{F}$.  

For the substrate, which is subjected to uniaxial tension along the X-direction, the stress conditions impose $T^s_{22} = T^s_{33} = 0$, leading to the following deformation gradient:  
\begin{equation}
    \mathbf{F}^{s} = \text{diag}\left[\lambda_{1}^s, (\lambda_{1}^s)^{-\nu_{s}}, (\lambda_{1}^s)^{-\nu_{s}}\right].
\label{eq:subsdef}
\end{equation}  

Because the film and substrate are perfectly bonded, they experience the same stretches in the X- and Z-directions, so that $\lambda_1^f = \lambda_1^s=\lambda_1$, $\lambda_3^f = \lambda_3^s = (\lambda_{1}^s)^{-\nu_{s}}$. Furthermore, the transverse stress component satisfies $T^f_{22} = T^s_{22} = 0$. Applying these conditions, we find the deformation gradient for the film as:
\begin{equation}
    \mathbf{F}^{f} = \text{diag}\left[\lambda_{1}, \lambda_{1}^{\frac{\nu_{f}(\nu_{s}-1)}{(1-\nu_{f})}}, \lambda_{1}^{-\nu_{s}}\right].
\label{eq:filmdef}
\end{equation}  
\crefrange{eq:subsdef}{eq:filmdef}, together with the stress expression in \cref{eq:cauchystress1}, define the base state (elastic solution) of the bilayer under tension.  

To determine the onset of wrinkling, a small-amplitude displacement $\mathbf{u}$ is superimposed on the finite deformations given in \crefrange{eq:subsdef}{eq:filmdef}. Following the incremental approach outlined by \citet{haughton1978incr}, the resulting incremental equilibrium equations are derived as follows:  
\begin{align}
    & \pazocal{A}_{0jilm}^{{s}} u_{m,lj}^{{s}}=0, &&  -\infty<y<0, \notag \\
& \pazocal{A}_{0jilm}^{{f}} u_{m,lj}^{{f}}=0, && 0<y<h,
\label{eq:equil}
\end{align}
where the commas denote differentiation with respect to the current coordinates, $h$ is the current thickness of the film, and $\pazocal{A}_{0}$ is the fourth-order tensor of the instantaneous elastic moduli, with components:
\begin{equation}
\pazocal{A}_{0jilm} = J^{-1} F_{j \alpha} \frac{\partial^2 W}{\partial F_{i \alpha} \partial F_{m \beta}} F_{l \beta}.
\label{eq:eqincrem}
\end{equation}

We look for solutions in the forms:
\begin{equation}
u_{z} = e^{\gamma y} \sin(kz), \qquad 
u_{y} = e^{\gamma y} \cos(kz),
\label{eq:trial}
\end{equation}
where $\gamma$ is the attenuation coefficient and $k$ is the wavenumber of the sinusoidal wrinkles.
Substitution into \cref{eq:equil} leads to an eigenproblem, with characteristic equation a bicubic in $\gamma$.
In the film, the general solution is of the form: 
\begin{equation}
\begin{split}
u_{z}^{{f}} &= \Big(\sum_{i=1}^{4} \pazocal{V}_{i}e^{\gamma_{i}y}\Big)\sin(kz), \quad 
u_{y}^{{f}} = \Big(\sum_{i=1}^{4} \pazocal{V}_{i}e^{\gamma_{i}y}\Big)\cos(kz),
\end{split}
\label{eq:wrinklfilm}
\end{equation}
where $\gamma_{1}, \cdots, \gamma_{4}$ are the eigenvalues, and $\pazocal{V}_{1}, \cdots, \pazocal{V}_{4}$ are constants. 

In the substrate, the stretch ratios along the Y and Z directions are equal, and $\gamma=1$ is a repeated eigenvalue. The other repeated root, $\gamma=-1$ is discarded to enforce decay. Therefore, the solution is of the form:
\begin{equation}
\begin{split}
u_{z}^{{s}} &= (\pazocal{U}_{1}+\pazocal{U}_{2}y)e^{y}\sin(kz), \qquad 
u_{y}^{{s}} = (\pazocal{U}_{1}+\pazocal{U}_{2}y)e^{y}\cos(kz),
\end{split}
\label{eq:wrinklsub}
\end{equation}
where $\pazocal{U}_{1}, \pazocal{U}_{2}$ are constants. \\

By applying the traction-free boundary conditions:
\begin{subequations}
\begin{equation}
\pazocal{A}_{02ilm}^{{f}} u_{m,l}^{{f}}=0,\qquad y=h,
\label{eq:tractionb}
\end{equation}
and the continuity conditions:
\begin{equation}
\pazocal{A}_{02ilm}^{{f}} u_{m,l}^{{f}}=\pazocal{A}_{02ilm}^{{s}} u_{m,l}^{{s}},\qquad u_{i}^{{f}}=u_{i}^{{s}}, \qquad y=0,
\label{eq:continuityb}
 \end{equation}
\end{subequations}
we obtain six homogeneous equations for $\lbrace{\pazocal{U}_{1}, \pazocal{U}_{2}, \pazocal{V}_{1}, \cdots, \pazocal{V}_{4}\rbrace}$. The bifurcation condition is then given by equating the determinant of a $6\times6$ coefficient matrix to 0. This condition provides the foundation for an asymptotic analysis.


\subsection{Asymptotic analysis}


Prior studies have predominantly examined the wrinkling of elastic bilayers subjected to uniaxial tension in the context of linear elasticity. 
To derive expressions for the critical strain $\varepsilon_c$ and critical wavenumber $k_c$ \citep{biot1937bending, timoshenko1959theory}, a two-step approach was adopted.

First, the critical compressive strain for a linear elastic bilayer system is derived by imposing the plane-strain condition. This derivation follows the classical ordinary differential equation governing the bending of a thin elastic beam, modeled under plane-stress conditions, on a semi-infinite elastic foundation subjected to a compressive load \citep{allen1969chapter, Cai2000}.
Second, the difference in lateral contraction between the substrate and film layers is used to approximate the critical strain under applied tensile loading. 

This approach resulted in the following expressions, derived by \citet{Volynskii2000}, \citet{ChungJY2011}, and \citet{Siavash2019}:
\begin{equation}
\varepsilon_c = \frac{1}{4(\nu_s - \nu_f)}\left(3\frac{\mu_{s}}{\mu_{f}}\frac{1-\nu_f}{1-\nu_s}\right)^{2 / 3}, \qquad
 k_ch =\left(3\frac{\mu_{s}}{\mu_{f}}\frac{1-\nu_f}{1-\nu_s}\right)^{1/3},\label{eq:nikra1}
\end{equation}
for linearly elastic bilayers under uniaxial tension. 
These expressions are used in buckling-based metrology applications to calculate the Young modulus of the film, but they are only valid for low strains ($\ll 10\%$) and under the assumption of plane strain \citep{Stafford2004}. 

However, it is important to note that the plane-strain assumption used in the derivation of \cref{eq:nikra1} is not valid for a bilayer system under uniaxial tension.
To address this gap, we follow \citet{Cai2000, Cai2019} and assume that the shear modulus ratio $r = {\mu_s}/{\mu_f}$ is of the order $(k_ch)^3$ to derive asymptotic expressions for the critical strain and wavenumber. For simplicity, we set $\alpha_s = \alpha_f = 1$ in \cref{eq:energy1} and focus on the compressible neo-Hookean strain energy function for both the film and substrate layers:
\begin{equation}
W^{\text{nH}} = \frac{\mu}{2}\left[I_1-3+\frac{1-2\nu}{\nu}\left(I_3^{-\frac{\nu}{1-2\nu}}-1\right)\right].
\label{eq:nH1}
\end{equation}

To derive the asymptotic expressions for the critical strain and wavenumber, we introduce the following ansatz:
\begin{equation}
\begin{split}
k_{c}h &= \sum_{i=1}^{5}d_{i}\delta^{i} + \pazocal{O}(\delta^{6}), \quad \lambda_{c} = 1+ \sum_{\substack{i=1 \\ i \ne 2}}^{6}c_{i}\delta^{i} + \pazocal{O}(\delta^{7}), \quad \delta = \left(\frac{\mu_{s}}{\mu_{f}}\right)^{1/3} = r^{1/3}.
\end{split}
\end{equation}
From \cref{eq:filmdef}, the component $F^f_{22}$ can be expanded as a series in terms of $\delta$, as: 
\begin{equation}
F_{22}^{f} = \lambda_{1}^{\frac{\nu_{f}(\nu_s-1)}{(1-\nu_{f})}} = 1+\sum_{i=1}^{8} \hat{d}_{i}\delta^{i} + \pazocal{O}(\delta^{9}).
\end{equation}
Similarly, the attenuation coefficients for the film layer $\gamma_{1}, \dots, \gamma_{4}$ and their squares can be expressed as series expansions in terms of \(\delta\) up to eight terms. These approximations significantly reduce computational complexity.

Next, we substitute the expressions for the attenuation coefficients ($\gamma_{1}, \dots, \gamma_{4}$), $k_c h$, $\lambda_{c}$, and $F_{22}^{f}$ into the bifurcation equation and its derivative with respect to $k_c h$, then expand both in terms of $\delta$. These expansions are truncated to 12 and 11 terms, respectively, with higher-order terms omitted for simplicity.

Finally, to determine the unknown parameters, we impose the condition that the coefficients of \( \delta^i \) and \( \delta^{i-1} \) in the bifurcation equation and its derivative (for all \( i \geq 8 \)) vanish simultaneously.  
This yields recursive relations for \( c_{j+1} \) and \( d_j \) for \( j \geq 1 \).

Using this procedure, we obtain the asymptotic expressions for the critical stretch and wavenumber:
\begin{align}
& \lambda_{c}  =  1   + \frac{\left[(1-\nu_f)(1-\nu_s)\right]^{2/3}}{(\nu_s - \nu_f)\left(8\nu_s/3-2\right)^{2/3}} r^{2/3} 
  + \frac{(1 - \nu_f)(2\nu_s - 1)}{(\nu_s - \nu_f)(4 \nu_s - 3)} r  + c_{4} r^{4/3} + c_{5} r^{5/3} 
 + c_{6} r^{2} + \pazocal{O}(r^{7/3}),
\notag \\
& k_c h  = \left[\frac{4(1-\nu_f)(1 - \nu_s)}{1 - 4\nu_s/3}\right]^{1/3} r^{1/3} + d_{3} r + d_{4} r^{4/3} + d_{5} r^{5/3} + \pazocal{O}(r^2),
\label{eq:nHlam}
\end{align}
where the coefficients $c_{4}, \dots, c_{6}, d_3, \dots, d_5$ are provided in \cref{asym}.

\addition{To provide insight into the wavelength of wrinkles, we derive scaling laws for bilayers and compare them with those observed in free-standing films. Using the leading-order term of critical wavenumber \cref{eq:nHlam}, we see that the wavelength of wrinkles ($l_w$) in bilayers scales as:}
\begin{subequations}
\begin{equation}
    l_w \propto h \left(\frac{1 - 4\nu_s/3}{(1 - \nu_f)(1-\nu_s)}\right)^{1/3},
\end{equation}
\addition{whereas for free-standing films, it scales as \citep{Cerda2003}:}
\begin{equation}
    l_w \propto h^{1/2} \left(1 - \nu_f^2\right)^{-1/4}.
\end{equation}
\label{eq:scalelaws}
\end{subequations}

\addition{To draw a direct comparison, we fix $\nu_s$ at any value between -1 and 0.5, so that \cref{eq:scalelaws}a gives the scaling $l_w \propto h\left(1 - \nu_f\right)^{-1/3}$. 
Recalling that the film thickness is small $(h \ll 1)$, and comparing with \cref{eq:scalelaws}b shows that the wavelength of wrinkles in bilayers is always shorter than that of free-standing films, highlighting the influence of the substrate on the wrinkle characteristics.}

\addition{These theoretical predictions are consistent with experimental observations: for bilayers, the wrinkle wavelength, is typically in the micro to nanoscale range, in line with the findings of \citet{Volynskii2000}, \citet{Stafford2004}, \citet{Yin2012}, \citet{Jin2015}, and \citet{Hu2016}, whereas for free-standing films, it is in the millimeter to micrometer scale according to the experimental results reported by \citet{cerda2002wrinkling}.}


\subsection{Homogenization and inverse analysis}
\label{inversedes}


Within a small-strain framework, the homogenized elastic stiffness $\overline{\mathbb{C}}$ is defined as
\begin{equation}
\label{eq:homC}
    \overline{\mathbb{C}} = \frac{\partial\overline{\boldsymbol{T}}}{\partial\overline{\boldsymbol{\varepsilon}}},
\end{equation}
where $\overline{\boldsymbol{T}}$ and $\overline{\boldsymbol{\varepsilon}}$ denote the homogenized Cauchy stress and homogenized small-strain tensors, respectively. These quantities are obtained by averaging the local stress $\boldsymbol{T}$ and strain $\boldsymbol{\varepsilon}$ over the volume of the unit cell $\Omega$:
\begin{equation}
\label{eq:homT}
    \overline{\boldsymbol{T}} = \frac{1}{\Omega} \int_{\Omega} {\boldsymbol{T}} \, \mathrm{d}\Omega, \qquad 
    \overline{\boldsymbol{\varepsilon}} = \frac{1}{\Omega} \int_{\Omega} {\boldsymbol{\varepsilon}} \, \mathrm{d}\Omega.
\end{equation}

The homogenized stiffness in \cref{eq:homC} can be computed using various approaches, such as FE-based periodic homogenization \citep{hassani1998review} or FFT-based techniques \citep{moulinec1998numerical}. 
However, in inverse analysis, the iterative nature of the inverse design procedure makes it impractical to directly use these homogenization schemes. 
An alternative approach is to construct machine learning surrogate models using datasets generated from the aforementioned physics-based methods, thereby enabling the inverse design process. 

\cite{danesh2024fft} adopted this latter approach for the inverse design of auxetic unit cells, where, first, the forward surrogate models $\mathcal{S}$ are generated to predict the homogenized elastic stiffness constants $\overline{C}_{ijkl} = \mathcal{S}(\mathbf{g},\mathbf{m})$ as functions of the vectors $\mathbf{g}$ and $\mathbf{m}$, denoting the geometric and material input parameters, respectively. Then, the inverse problem is solved to obtain the unit cell's geometric parameters by minimizing the loss between the target homogenized stiffness component $\hat{\overline{C}}_{ijkl}$ and its counterpart predicted by the surrogate model $\overline{C}_{ijkl}$.
The loss function is defined as a weighted sum of squared differences between the predicted and target homogenized stiffness components:
\begin{equation}
\label{eq:lossinv}
    \mathcal{L}_{\mathrm{inv}} = \frac{1}{n_c} \sum_{m=1}^{n_{\text{ind}}} w_{m}\left(\overline{C}_{ijkl}-\hat{\overline{C}}_{ijkl}\right)^2.
\end{equation}
The index $m$ enumerates the $n_{\text{ind}}$ independent components of the stiffness tensor $\overline{\mathbb{C}}$. Each independent component $\overline{C}_{ijkl}$ corresponds to a specific $m$, and the Boolean variable $w_m$ determines whether that particular component is included in the loss calculation ($w_m = 1$) or ignored ($w_m = 0$). The normalization factor $n_c = \sum_{m=1}^{n_{\text{ind}}} w_m$ ensures that the loss is averaged over only the selected components. Although the summation is formally written over $m$, it implicitly accounts for all relevant $C_{ijkl}$ components through their indexing. For orthogonal void auxetic unit cells, the presence of symmetry axes results in orthotropic stiffness characterized by three independent elastic constants (i.e. $n_{\text{ind}}=3$). 

Minimizing the loss function in \cref{eq:lossinv} using gradient-based optimization \citep{zheng2021data} or grid search algorithms \citep{danesh2024fft} yields the optimal geometric parameters that achieve the desired homogenized stiffness. 
We use this approach in \cref{Res_Hom} to determine the geometric parameters of auxetic unit cells with the desired Poisson ratios.


\section{Finite element simulations: 3D models}
\label{methods}


This section presents our FE approach using 3D models to study wrinkling in homogenized bilayers and those with microstructured films of different material properties. Initially, we establish and validate our numerical setup. We then explore two case studies: one that examines bilayers with incompressible substrates and the other with highly auxetic films. Our analysis concludes with predictions from asymptotic expressions and the application of inverse analysis techniques, followed by a comparison of the buckling behavior of bilayers with microstructured films to predictions from homogenized models.


\subsection{Numerical setup and validation}\label{method2}


To validate the accuracy of our implementation of periodic boundary conditions (PBCs) using a Python script, we conducted a preliminary study using a linear buckling analysis for an incompressible neo-Hookean bilayer under 2D (plane-strain) compression. 
The bilayer was periodic along the X-direction (see \cref{2Dmodel}), and we successfully recovered the results reported by \citet{Cao2012b}, confirming the reliability of our approach.

Following this validation, we performed a series of FE simulations to investigate wrinkling in 3D bilayers subjected to uniaxial tension. The bilayers were modeled using the Blatz--Ko strain energy function, considering two primary configurations: one with an incompressible substrate and the other with a highly auxetic film.

The depth of the bilayer (in the Z-direction) was set as an integer multiple of the predicted wrinkling wavelength, calculated using semi-analytical results from \textsf{Mathematica}, for the cases: $(\nu_{f}, \nu_{s})=\{0.3, 0.495\}$ and $\{-0.95, -0.8\}$. For other pairs of Poisson ratios, small variations in depth were observed to have a negligible effect on the critical stretch values, with changes only in the third decimal place. 

To optimize computational efficiency without compromising accuracy, we used the following dimensions. 
For the incompressible substrate case, the bilayer structure had dimensions (X, Y, Z) of \( 15 \times 45 \times 5.684 \) units, with a thickness ratio of \( h_s/h_f = 299 \) (\cref{3dmodel2a}). In the case of the highly auxetic film, the dimensions were \( 15 \times 80 \times 4.504 \) units, with a thickness ratio of \( h_s/h_f = 399 \) (\cref{3dmodel2b}). For the neo-Hookean bilayers ($\alpha_f=\alpha_s=1$), the dimensions were \( 6 \times 45 \times 1.85 \) units, with a thickness ratio of \( h_s/h_f = 299 \) (\cref{3dmodel2b1}).


\begin{figure}[!ht]
\centering
\includegraphics[width=\textwidth]{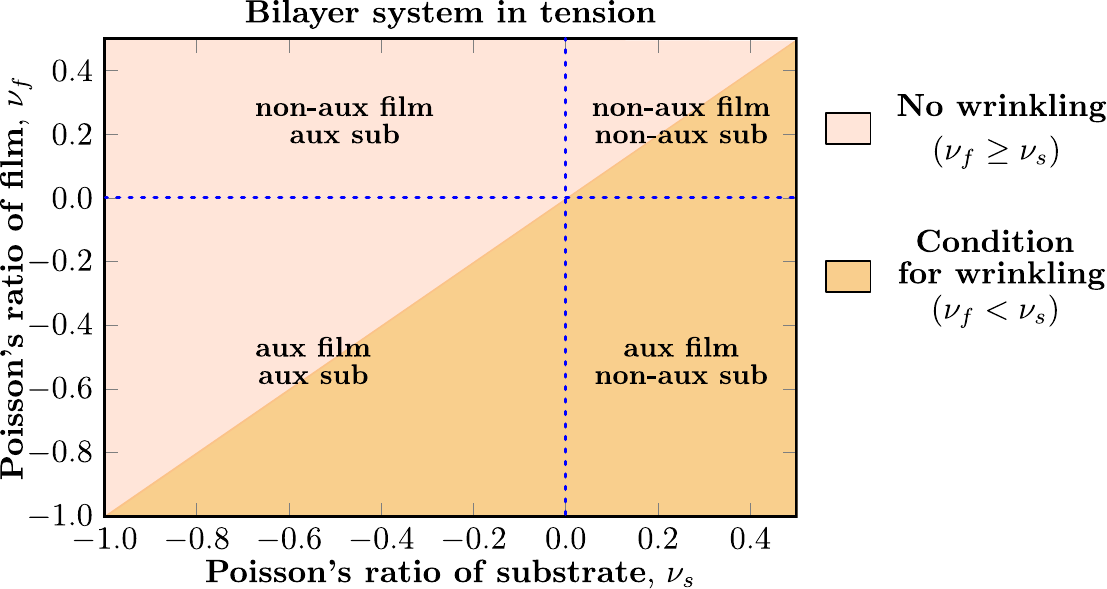}
    \caption{Wrinkling condition according to the contrast between the Poisson ratios of film and substrate, $\nu_f$ and $\nu_s$, respectively.}
    \label{fig:wrinklcond}
\end{figure}


PBCs were applied on the extreme faces in the X- and Z-directions of the domain using a user-defined Python script. Both the film and the substrate layers were modeled using a \textsf{UHYPER} subroutine with the Blatz--Ko strain energy function.

FE discretization was performed using 20-node hexahedral elements with quadratic interpolation and reduced integration (\textsf{C3D20R}) for both the film and substrate layers. For the limiting case of incompressibility, \textsf{C3D20RH} elements were used.

For the Blatz--Ko bilayers, the FE mesh consisted of 34,200 elements for the incompressible substrate case (\cref{3dmodel2a}) and 17,130 elements for the highly auxetic film case (\cref{3dmodel2b}). 

\addition{\citet{Cai1999} demonstrated that the stability of a coated elastic half-space depends on the shear modulus ratio $r = \mu_s / \mu_f$. The system becomes less stable when $r < 1$ and more stable when $r > 1$. Physically, when the film is stiffer than the substrate ($r < 1$), it carries the majority of the load, resulting in the buckling of the film layer as an energetically favorable state. This occurs because the system minimizes its total energy by allowing the film to buckle rather than undergo uniform compression. In contrast, when the film is softer than the substrate ($r > 1$), the stability of the half-space increases, suppressing wrinkling and favoring homogeneous in-plane deformation.}

\addition{For neo-Hookean bilayers, \citet{Cao2012b} observed that when the film is sufficiently stiff ($r < 0.1$), sinusoidal wrinkling remains the stable and preferred mode within the linear buckling regime. However, when the film is relatively soft ($r > 0.5$), surface creases (nonlinear buckling) can emerge within the film before the wrinkling bifurcation mode, as noted by \citet{Hong2009}.}

Based on these observations, we set the Blatz–Ko parameters to $\alpha_f = \alpha_s = 0.4$ and the shear modulus ratio to $\mu_f/\mu_s=30$ or $r=1/30$ for FE simulations. For 3D bilayer systems under large uniaxial tension, we found that wrinkles were observed only when the Poisson ratio of the film ($\nu_f$) was lower than that of the substrate ($\nu_s$), as summarized in \cref{fig:wrinklcond}. This result is consistent with the findings of \citet{Siavash2019}, which were obtained for linear-elastic bilayer systems.



Following \cref{fig:wrinklcond}, we chose two cases detailed in \crefrange{3dmodel2a}{3dmodel2b}, based on the Poisson ratio of the layers, to study wrinkling in the layered system under uniaxial tension. We find that the numerical results obtained using different approaches in the software \textsf{ABAQUS} and \textsf{Mathematica} match well for a wide range of material parameters.


\subsection{Incompressible substrate}
\label{3dmodel2a}



\begin{figure*}[!ht]
\centering
\includegraphics[width=\textwidth]{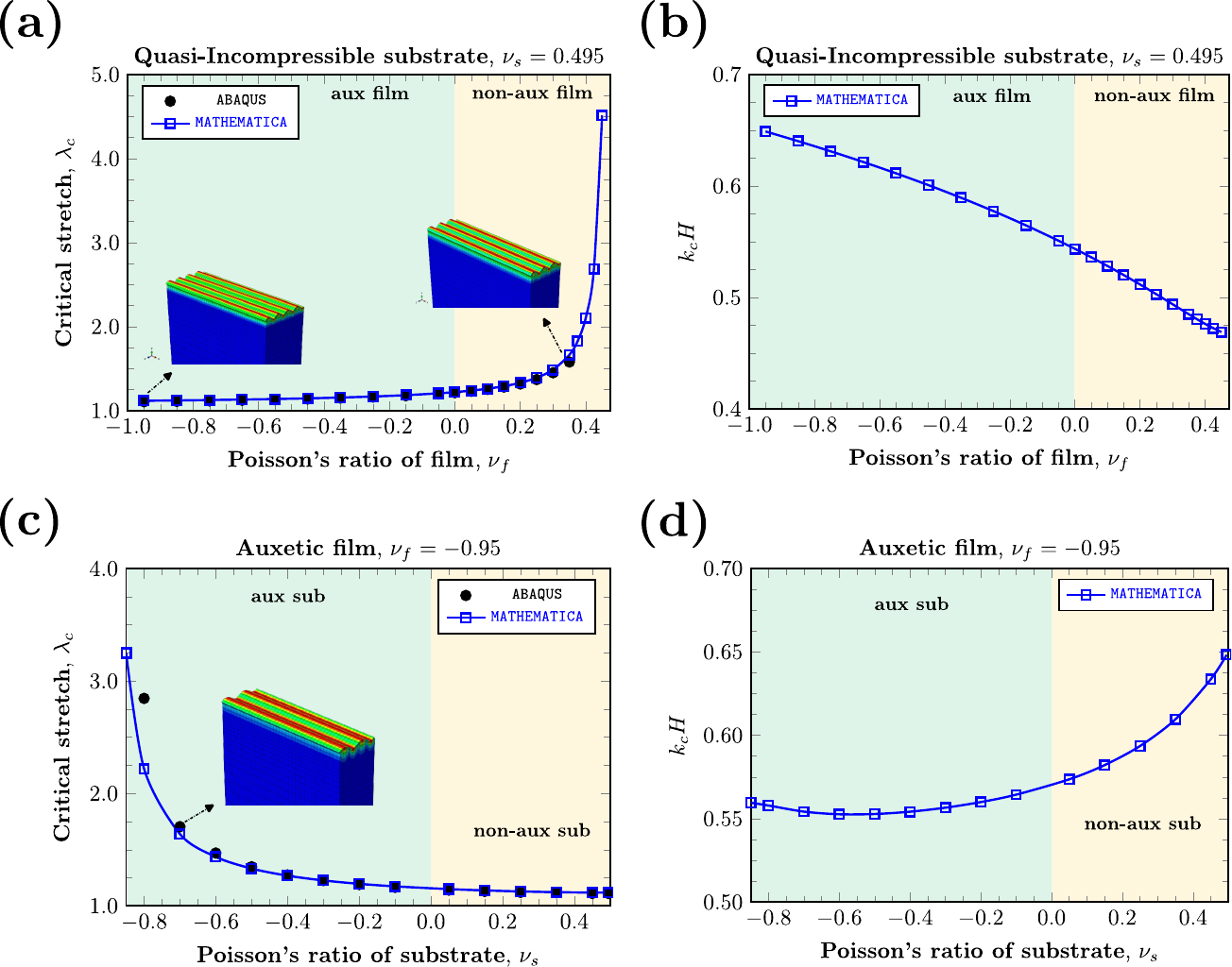}
    \caption{Numerical simulations using Blatz--Ko material model (\cref{eq:energy1}). Variations of the critical stretch of wrinkling $\lambda_c$  and corresponding critical wavenumber measure $k_{c}H$ with the Poisson ratio of one layer ($k_c$ is the critical wavenumber and $H$ is the film initial thickness). \addition{We set the shear modulus ratio to $\mu_f/\mu_s=30$, i.e. $r=1/30$}. (a-b): The substrate is quasi-incompressible ($\nu_s = 0.495$). (c-d): The film is highly auxetic ($\nu_f = -0.95$). Results from \textsf{ABAQUS}: black dots, results from \textsf{Mathematica}: solid line with square markers.}
    \label{fig:3dwrinklingabamath}
\end{figure*}


First, we assume that the substrate is quasi-incompressible ($\nu_s = 0.495$), while the film's Poisson ratio varies within the range $\nu_f \in [-0.95, 0.495)$. 
When the substrate has a higher Poisson ratio than the film, it undergoes greater transverse compression (Z-direction) under uniaxial tension applied along the X-direction. This mismatch in deformation between the layers generates compressive stresses in the film, leading to wrinkle formation parallel to the X-direction (see \cref{fig:3dmodel}).

Variations in the critical stretch for wrinkling $\lambda_c$ and the corresponding critical wavenumber $k_c$ are plotted against $\nu_f$ in \cref{fig:3dwrinklingabamath}(a-b). 
As shown in \cref{fig:3dwrinklingabamath}(a), the critical stretch values predicted by the semi-analytical analysis (\textsf{Mathematica}) and the FE buckling analysis (\textsf{ABAQUS}) align well for $\nu_f \leq 0.35$. 
However, beyond this threshold, \textsf{ABAQUS} ceases to predict wrinkles in the desired direction and instead returns negative eigenvalues. This suggests that the load direction must be reversed to obtain wrinkles, which is unphysical and indicates a breakdown of the numerical predictions.  

For an auxetic film ($\nu_f < 0$), expansion occurs in all directions under tension, while the incompressible substrate contracts along the Z-direction, generating compressive stress in the film. Consequently, wrinkles form at lower critical stretch values. As the Poisson ratio of the film approaches that of the substrate, the critical stretch required for wrinkling increases sharply, consistent with the findings in \cref{fig:wrinklcond}.


\subsection{Highly auxetic film}
\label{3dmodel2b}


To examine the effect of a highly auxetic film, we consider a film layer with a negative Poisson ratio, $\nu_f = -0.95$, while varying the Poisson ratio of the substrate within the range $\nu_s \in (-0.95, 0.495]$.  

\cref{fig:3dwrinklingabamath}(c-d) displays the variations in the critical stretch $\lambda_c$ and the wavenumber $k_c$ as a function of $\nu_s$. Our numerical results from \textsf{ABAQUS} and semi-analytical predictions from \textsf{Mathematica} show strong agreement in the range $-0.7\leq\nu_s\leq0.495$. 
However, for $\nu_s < -0.8$, \textsf{ABAQUS} fails to provide meaningful predictions.  

The results indicate that buckling generally occurs at lower stretch values for highly auxetic films. 
However, when $\nu_s$ approaches $\nu_f$, the film and substrate exhibit similar levels of transverse contraction. 
As a result, the compressive stress developed in the film layer is insufficient to induce wrinkle formation.

\addition{Typically, the wavelength of wrinkles increases as the Poisson ratios of the film and substrate approach each other and decreases as they diverge, as shown in \cref{fig:3dwrinklingabamath}(b,d). However, when the film and substrate are highly compressible $(\nu_{f} = -0.95 \hspace{1mm} \text{and} \hspace{1mm} \nu_{s} \leq -0.7)$, this trend reverses, and the wavelength decreases, as illustrated in \cref{fig:3dwrinklingabamath}(d).}

\addition{Notably, this behavior is not observed in compressible neo-Hookean bilayers (\cref{eq:nH1}). This can be verified by plotting the asymptotic expression for the critical wavenumber (\cref{eq:nHlam}), which remains a monotonically decreasing function as the Poisson’s ratios of the film and substrate approach each other. The distinct behavior observed in Blatz–Ko-type materials for $\nu_{s} \leq -0.7$ deviates from expected trends and can be further explored in future work.}

In summary, we found that to ensure early wrinkling in a bilayer subjected to uniaxial tension, a large contrast in the Poisson ratio between the film and the substrate is required. This effect is particularly pronounced when one material is auxetic and the other is non-auxetic, highlighting the critical role played by the Poisson ratio contrast in governing instabilities in bilayer systems.


\subsection{Asymptotic solution: Compressible bilayer system}
\label{3dmodel2b1}



\begin{figure*}[!ht]
\centering
\includegraphics[width=\textwidth]{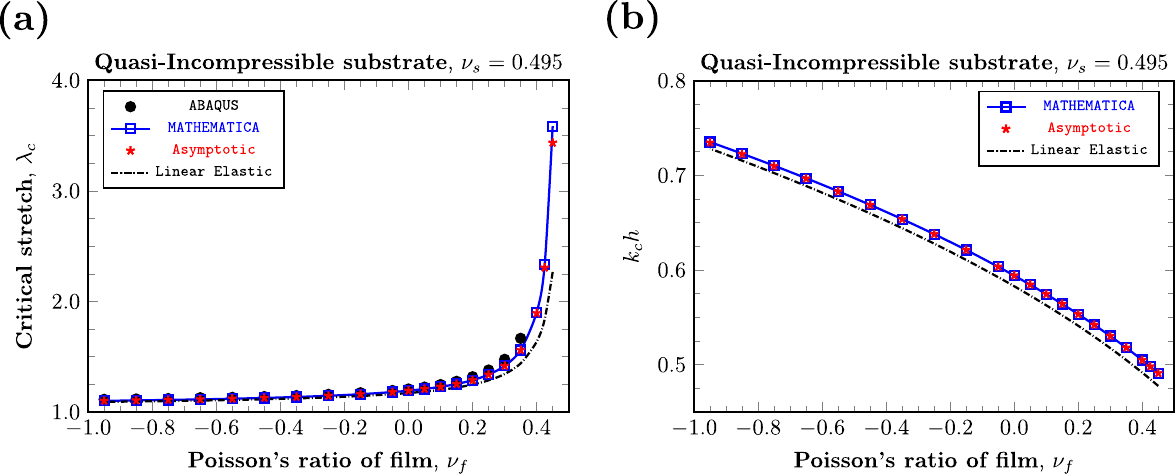}
    \caption{Numerical simulations for a bilayer made of compressible neo-Hookean materials ($\alpha_s = \alpha_f = 1$ in \cref{eq:energy1}). The substrate is quasi-incompressible ($\nu_s = 0.495$). \addition{We set the shear modulus ratio to $\mu_f/\mu_s=30$, i.e. $r=1/30$}. (a-b): Variations of the critical stretch of wrinkling $\lambda_c$  and corresponding critical wavenumber measure $k_{c}h$ ($k_c$: critical wavenumber, $h$: current film thickness) with the Poisson ratio of film layer $\nu_f$.  \textsf{ABAQUS}: black dots, \textsf{Mathematica}: solid line with square markers, \textsf{Asymptotic} expressions (\cref{eq:nHlam}): star markers, \textsf{Linear-elastic} expressions (\cref{eq:nikra1}): dash-dotted line.}
    \label{fig:nHstretchwavenumber}
\end{figure*}



\begin{figure*}[btp]
\centering
\includegraphics[width=\textwidth]{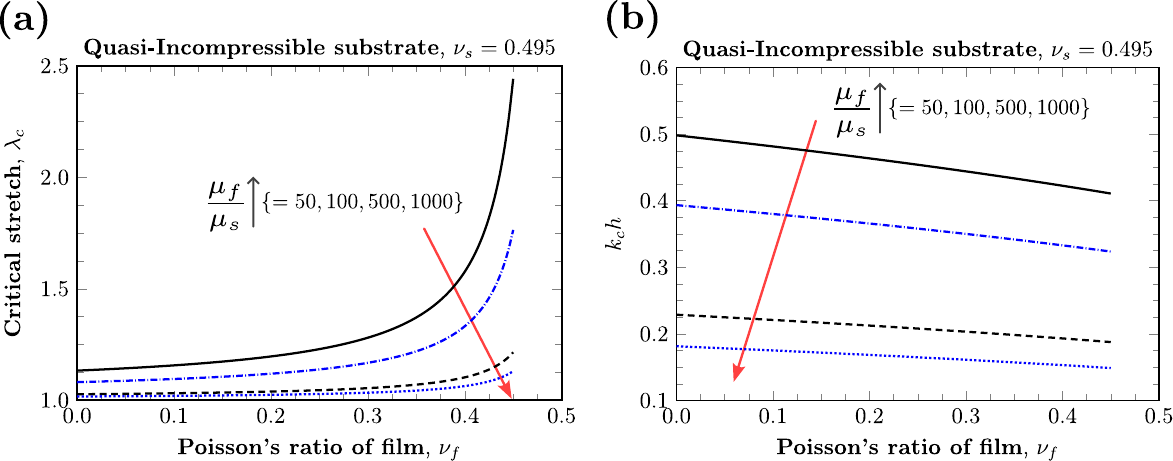}
    \caption{Using the asymptotic expressions \cref{eq:nHlam} for compressible neo-Hookean materials ($\alpha_s = \alpha_f = 1$ in \cref{eq:energy1}). 
    The substrate is quasi-incompressible ($\nu_s = 0.495$). (a-b): Variations of the critical stretch of wrinkling $\lambda_c$  and corresponding critical wavenumber measure $k_{c}h$ with the Poisson ratio of film layer ($k_c$ is the critical wavenumber and $h$ is the deformed film thickness) and for contrast in shear moduli between the layers $(\mu_{f}/\mu_{s} = \lbrace 50, 100, 500, 1000\rbrace)$.}
    \label{fig:nHvaryr}
\end{figure*}


To further analyze the influence of material nonlinearity and Poisson ratio contrast, we compare the predictions from linear (\cref{eq:nikra1}) and hyperelastic models (\cref{eq:nHlam}).

When strains are small, the asymptotic expressions for compressible neo-Hookean bilayers (\cref{eq:nH1}) recover the predictions of the linear elastic model. However, as $\nu_f$ approaches $\nu_s$, the critical strain increases significantly and the linear elastic model (\cref{eq:nikra1}) fails to capture accurate behavior due to the breakdown of the plane-strain assumption and the effects of non-linearities in geometric and material parameters. Consequently, linear elastic expressions substantially underestimate numerical results from \textsf{ABAQUS} and \textsf{Mathematica}, whereas the hyperelastic model remains in good agreement.

Meanwhile, when the shear modulus ratio ($\mu_f/\mu_s$) and the Poisson ratio contrast ($\nu_s - \nu_f$) are high, the plane-strain condition remains approximately valid as $\lambda_c \to 1$. In this regime, the linear elastic model aligns well with numerical results and the asymptotic solution. This behavior is illustrated in \cref{fig:nHstretchwavenumber}(a), which shows the variation of $\lambda_c$ with $\nu_f$ for a quasi-incompressible substrate ($\nu_s = 0.495$).

In \cref{fig:nHstretchwavenumber}(b), although a similar trend is observed in both linear and nonlinear models, we see that the linear elastic model consistently underpredicts the critical wavenumber compared to numerical and asymptotic results. In particular, when the Poisson ratio contrast is high (low), wrinkles exhibit shorter (longer) wavelengths, as seen in \cref{fig:3dwrinklingabamath}.  

\cref{fig:nHvaryr} illustrates the dependence of $\lambda_c$ and $k_ch$ on $\nu_f$ for different shear modulus contrasts $\mu_f/\mu_s$. According to \cref{eq:nHlam}, with $r$ assumed to be on the order of $(k_ch)^3$, an increase in $\mu_f/\mu_s$ or a decrease in $r$ results in a lower critical stretch and wavenumber, leading to early buckling and long-wavelength wrinkles.

For compressible neo-Hookean bilayers, the critical stretch ratio for buckling and the wrinkle wavelength increase as the Poisson ratios of the film and substrate approach each other. Asymptotic expressions generate curves instantly, offering a computationally efficient alternative to the \textsf{ABAQUS} simulations.  Building on these findings, we extend our analysis to homogenization and inverse techniques, with the aim of tailoring auxetic microstructures to achieve desired material properties.


\subsection{Inverse design and homogenization: Application}
\label{Res_Hom}



\begin{figure*}[htb]
\center
\includegraphics[width=\textwidth]{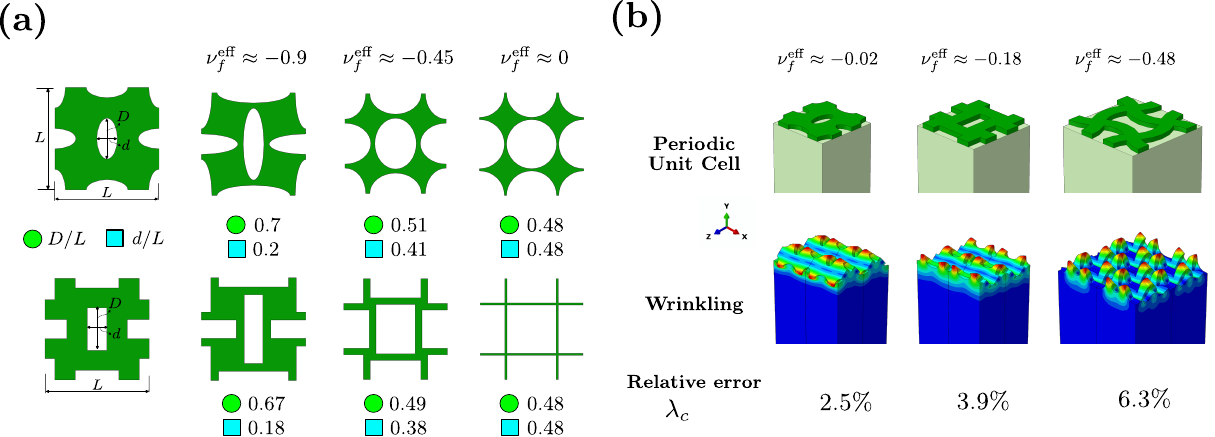}
    \caption{(a) Influence of microstructural geometry on the effective Poisson's ratio ($\nu_{f}^{\text{eff}}$) in auxetic films with orthogonal-oval and rectangular-void patterns. Films with desired effective Poisson's ratios are obtained using inverse analysis. (b) Comparison of undeformed and deformed configurations in bilayers with distinct auxetic microstructures, each designed to achieve specific Poisson's ratio in the small strain regime. Film and substrate base materials exhibit compressible ($\nu_f = 0.1$) and nearly incompressible ($\nu_f = 0.495$) properties, respectively. Periodic boundary conditions are applied on the lateral faces of the domain along X and Z and the system is subjected to uni-axial tension along X. We observe that the wrinkles are generated parallel to the direction of tension. The bottom of (b) shows the percentage relative error in critical stretch between homogenized bilayers (\cref{fig:3dwrinklingabamath}(a)) and those with microstructural patterns, obtained in \textsf{ABAQUS} with the Blatz--Ko model.}
    \label{fig:fig7}
\end{figure*}


Based on the surrogate models developed by \citet{danesh2024fft} for orthogonal void auxetic structures in the small strain regime, we use an inverse design approach to efficiently determine the geometric parameters of a chosen auxetic unit cell. 
This method precisely tailors the unit cell to achieve specific components of effective elastic stiffness based on the material properties of the base constituent. 
Using the Voigt notation, \citet{danesh2024fft} demonstrated that the in-plane effective stress of these orthogonal void auxetic unit cells can be directly correlated with their effective strain through the components of the effective stiffness as
\begin{equation}
    \begin{pmatrix}
        \overline{T}_{11}\\
        \overline{T}_{22}\\
        \overline{T}_{12}
    \end{pmatrix} = 
    \begin{pmatrix}
        \overline{C}_{11} & \overline{C}_{12} & 0\\
        \overline{C}_{12} & \overline{C}_{11} & 0\\
        0 & 0 & \overline{C}_{33}
    \end{pmatrix}
    \begin{pmatrix}
        \overline\varepsilon_{11}\\
        \overline\varepsilon_{22}\\
        2\overline\varepsilon_{12}
    \end{pmatrix}.
\end{equation}

Here, the notation $\overline{(\bullet)}$ refers to the homogenized average value of a quantity, including stress, strain, and elastic stiffness components. Therefore, if we define the components of the effective stiffness $\overline{C}_{11}$ and $\overline{C}_{12}$, we can obtain the desired effective Poisson's ratio $\nu_{f}^{\text{eff}}=\overline{C}_{12}/\overline{C}_{11}$. 

We now demonstrate how our analysis can be used to predict wrinkling instabilities in bilayers with auxetic patterns. In fact, auxetic properties can be obtained at a continuum level with a careful design of holes or voids at the microscale \citep{lakes1993advances, bertoldi2017flexible}.

\cref{fig:fig7}(a) illustrates how the geometric parameters of orthogonal-oval and rectangular voids influence the effective Poisson ratio ($\nu_{f}^{\text{eff}}$). Using the inverse design approach described by \citet{danesh2024fft}, we designed a set of auxetic structures with the desired effective Poisson ratios ($\nu_{f}^{\text{eff}}$) in the small-strain regime. 

To validate the effective properties of these structures generated from the inverse design, we used the finite element code \textsf{FEAP} \citep{taylor2014feap}. For this purpose, we discretized the unit cells with linear quadrilateral elements and performed homogenization to compute the effective elastic stiffness. Periodic boundary conditions were applied by enforcing symmetric nodes on opposing boundaries. After computing the effective elastic stiffness of each unit cell, we confirmed that the designed unit cells accurately recovered the target effective Poisson ratios with high precision.

After understanding the effect of geometric parameters of voids on the effective Poisson ratio, we consider an isotropic material with base properties of Young’s modulus $E_f = 110$ GPa and Poisson’s ratio $\nu_f = 0.1$. 
Using the inverse design approach, we generated auxetic structures with targeted effective Poisson’s ratios. 
Specifically, we designed film layers with orthogonal-oval, rectangular, and sinusoidal voids to achieve $\nu_{f}^{\text{eff}} = -0.02, -0.18$, and $-0.48$, respectively, in the small-strain regime. 
These structures were then used to analyze wrinkling behavior, as shown in \cref{fig:fig7}(b).

The critical stretch ratio for buckling in bilayers with these microstructural patterns aligns closely with the homogenized model predictions presented in \cref{fig:3dwrinklingabamath}(a), with a maximum relative error of 6.3\%. This agreement demonstrates the accuracy of our numerical method in predicting wrinkling instabilities in bilayers with engineered microstructures.


\section{Conclusions}
\label{conclusions}


Our study explores the possibility of harnessing wrinkles parallel to the direction of applied tension in 3D isotropic compressible bilayers subject to finite deformations. Particular emphasis was placed on configurations where the substrate is quasi-incompressible and when the film is highly auxetic.

We adopted a semi-analytical approach using \textsf{Mathematica} to predict the onset of wrinkling. For the finite element aspect of our study, we developed a new \textsf{UHYPER} subroutine tailored for Blatz-Ko material models, along with custom Python scripts designed to simulate wrinkling in compressible bilayers under periodic boundary conditions. These scripts facilitate linear buckling analysis in finite element simulations with \textsf{ABAQUS}.

For compressible neo--Hookean bilayer systems, we derived asymptotic expressions for the critical stretch ratio and critical wavenumber, which can be used under finite strains to determine the Young modulus of the film layer for buckling-based metrology applications. 

We found that wrinkles can be obtained only when the Poisson ratio of the substrate is greater than that of the film. 
When the Poisson ratios of film and substrate converge to a common value, the critical stretch ratio shoots up sharply, and the wavelength of wrinkles is high.
In the limit, the wrinkles are not present because there are no compressive stresses developing when the lateral expansions are the same for the film and substrate. 
Through multiple simulations, we showed that we can harness or delay the onset of wrinkles by varying the material properties.

Furthermore, using inverse analysis, we designed film microstructures to achieve desired effective Poisson ratios, validated with the finite element code \textsf{FEAP}. The critical stretch ratio for buckling in auxetic structures with microstructural patterns aligns closely with predictions from homogenized models.

Some of the limitations of our work include the consideration of isotropic strain energy functions for auxetic structures and deformation-independent effective material properties. 
Functional-grading of auxetics could also be explored with the methods presented in this study, see some preliminary works on harnessing instabilities in functionally-graded auxetic materials using tension-field theory \citep{Venkata2023, venkata2024} and their 
applications \citep{zhao2018programmed, zahoor2020preliminary, babivc2021challenges}. 

Ultimately, the method developed in this work could play a critical role in the manufacturing and testing of auxetic hydrogel organ patches \citep{Nguyen2022} and skin grafts \citep{Gupta2023}. This approach could specifically guide the design of graft microstructures to prevent undesired wrinkling.

\appendix


\section*{Appendix A: Short recaps of literature on auxetics and wrinkling instability of layered media}
\refstepcounter{section}
\label{AppendixA}


\setcounter{equation}{0}
\setcounter{figure}{0}
\renewcommand{\theequation}{A.\arabic{equation}}
\renewcommand{\thefigure}{A.\arabic{figure}}

\textit{Auxetics}, materials with negative Poisson's ratio, expand in all directions under uniaxial tension. 
For 3D isotropic materials with auxetic behavior, which obey the pointwise energy stability criterion, the theoretical value of Poisson's ratio ranges between -1 and 0.5 \citep{timoshenko1983history}.

Advancements in additive and subtractive manufacturing \citep{Mueller2013, sun2017powder}, combined with extensive research on negative Poisson ratio materials, have bridged the gap between theoretical models and practical applications of auxetics in diverse fields. Early studies on ferromagnetic films identified a negative behavior of the Poisson ratio that decayed with time \citep{popereka1970ferromagnetic}, while subsequent work established its theoretical existence in crystalline structures \citep{milstein1979existence}. \citet{lakes1987foam} experimentally demonstrated the auxetic behavior of open-cell polyurethane foams with tailored microstructures using thermo-mechanical treatments, and \citet{wojciechowski1989negative} used molecular simulations to explore auxetic behavior in two-dimensional isotropic systems. Advances in material design have extended the use of auxetic properties polymer foams to wrestling mats and aircraft sandwich panels \citep{lakes1993advances}. Recent breakthroughs in biomedical engineering have introduced auxetic materials for medical implants, including negative Poisson ratio stents for low flexural rigidity and high circumferential strength \citep{Dolla2006}, meta-implants for improved load distribution in hip replacements \citep{kolken2018rationally}, and hydrogel-based organ patches that adapt to dynamic tissue mechanics \citep{Nguyen2022}.

\textit{Bilayers} composed of compliant substrates coated with thin stiff films are commonly found in nature; for example, skin tissue consists of a thin, stiff epidermal layer attached to a thick, soft dermis. When subject to mechanical loads, bilayer systems can exhibit surface patterns through wrinkles.  
This instability phenomenon has found a wide range of applications in optical sensors \citep{Bowden1998}, novel flexible electronics \citep{Wagner2004, Khang2006, Zhang2019}, tunable phase gratings \citep{Harrison2004}, buckling-based metrology \citep{Stafford2004, Huang2007}, surface wetting \citep{Lin2009}, and buckling-related applications in soft matter \citep{Crosby2010}.

Historically, instability analysis in elastic half-spaces gained prominence through the seminal works of \citet{Biot1957}, \citet{Biot1963} and \citet{allen1969}. \citet{Dorris1980} studied instabilities in bilayer systems with different material combinations of elastic and inelastic film-substrate layers. \citet{Shield1994} conducted a buckling analysis of a stiff layer bonded to an isotropic, elastic half-space using first-order perturbation analysis and compared the results to approximate solutions obtained via beam theory for the coating. Building on Gurtin--Murdoch theory \citep{Gurtin1975, Gurtin1978}, \citet{steigmann1997} developed an incremental bifurcation theory for planar deformations of coated elastic solids with boundary elasticity.

Numerous studies have examined compression-induced surface instabilities in bilayers, driven by thermo-mechanical treatments. Experiments revealed surface instabilities, where thermal contraction of the substrate caused buckling in metal films \citep{Bowden1998}. Similarly, in metal-elastomer bilayers, thermal expansion mismatch during cooling induced equi-biaxial compression, leading to herringbone wrinkles \citep{Chen2004}. Refined models demonstrated that wrinkles evolve into stripes, labyrinths, or herringbones, depending on the anisotropy of the membrane force \citep{Huang2005}. \citet{Choi2007} demonstrated that silicon nanomembranes on pre-strained elastomeric substrates form 2D wavy structures under uniaxial and biaxial strains, offering potential for stretchable electronics. \addition{\citet{auguste2014} reported that an applied pre-tension in the substrate leads to the stabilization of wrinkles, while a pre-compression leads to the emergence of chaotic surface morphologies in thin coating films.}

A finite-deformation buckling theory was developed to describe the strain-dependent wavelength of wrinkles by \citet{Song2008}. \citet{Hong2009} found that creasing in bent elastomers occurs at a lower critical strain than the Biot prediction, which aligns with experiments. \citet{huang2017study} showed that surface instabilities in 2D bilayers under compression depend on the Poisson ratio of the substrate, occurring at smaller strains in incompressible substrates than in auxetics. \addition{\citet{cutolo2020} experimentally investigated the evolution of wrinkles in thin metal films bonded to substrates with 3D periodic surface structures.}
Material anisotropy and loading biaxiality influence wrinkling evolution in orthotropic films bound to compliant substrates, with phase diagrams revealing various post-buckling surface patterns, including stripes, checkerboard, and herringbone modes \citep{Yin2018}.

The formation of these surface patterns is controlled by different parameters, such as the contrast in material properties \citep{Cai1999, Efimenko2005}, differential growth \citep{Goriely2005}, film-to-substrate thickness ratio \citep{Stafford2005, Li2019}, initial imperfections \citep{Cao2012a}, nonlinearity of the substrate \citep{Brau2013, Zhuo2015}, curvature \citep{stoop2015}, applied prestretch \citep{Cao2012b, Cai2019}, origin of compression \citep{andres2018}, interfacial mechanics \citep{Bigoni2018, derya2023}, etc.


\section*{Appendix B: Validation of PBCs code -- 2D incompressible neo-Hookean bilayer model}\refstepcounter{section}\label{2Dmodel}


\setcounter{equation}{0}
\setcounter{figure}{0}
\renewcommand{\theequation}{B.\arabic{equation}}
\renewcommand{\thefigure}{B.\arabic{figure}}


\begin{figure}[!ht]
\centering
\includegraphics[width=0.85\textwidth]{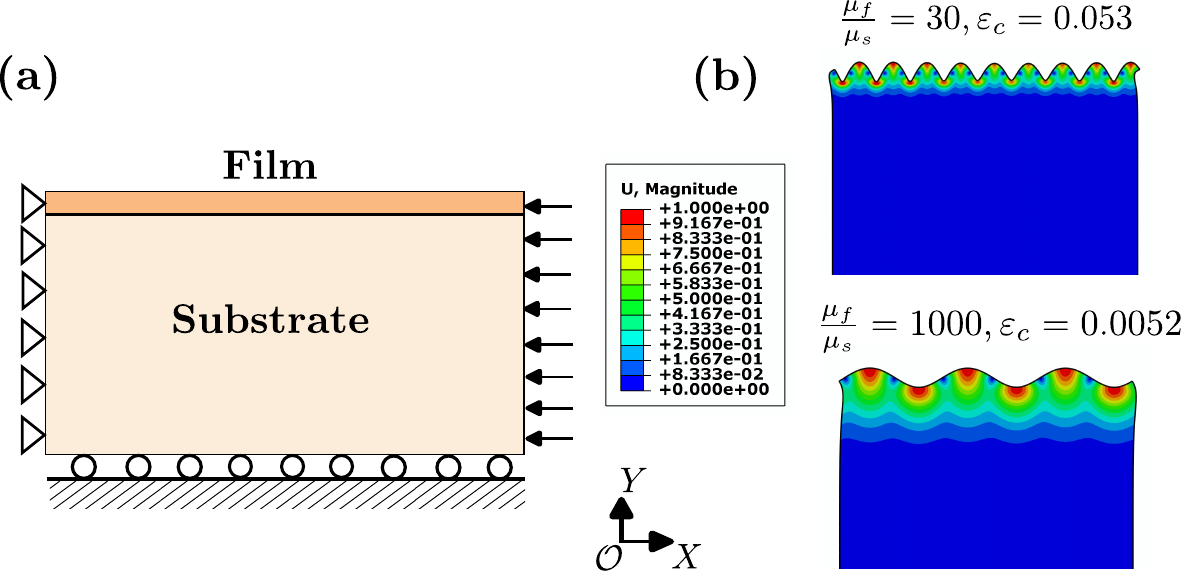}
    \caption{(a) Schematic representation of an incompressible neo-Hookean stiff film/soft substrate bilayer system under uni-axial compression. On the left edge of the domain, displacement and shear traction are set at zero. On the bottom edge of the domain, roller support restricts vertical displacement and shear traction, and the top surface of the film is traction-free. (b) Linear buckling solutions with critical strains $\varepsilon_c$ when ${\mu_f}/{\mu_s}=30, 1000$, in line with the predictions of \citet{Cao2012b}.}
    \label{fig:wrinkl2d}
\end{figure}

Under uniaxial, plane-strain compression, the theoretical critical strain is $\varepsilon_c=\nicefrac{1}{4}\left({3\mu_{s}}/{\mu_{f}}\right)^{\nicefrac{2}{3}}$ \citep{Cao2012b}.

In \cref{fig:wrinkl2d}, we applied PBCs on the left and right edges of the domain, roller support on the bottom edge of the substrate, perfect bonding between film and substrate layers, and traction-free condition on the top surface of the film in the two cases ${\mu_f}/{\mu_s}=30, 1000$. We find $\varepsilon_c = 0.053, 0.0052$, respectively, matching well with the theoretical solutions.

Having validated our 1D PBCs script along the X-direction on a 2D model, we then extended the PBCs to two dimensions, along the X- and Z-directions, to perform a linear buckling analysis on 3D bilayer systems under uniaxial tension.

The height ratio of substrate to film layers was taken as ${h_{{s}}}/{h_{{f}}}=163$, with the width of layers being 30.457 units. We modelled both film and substrate with the incompressible neo-Hookean model.

We used a hybrid 8-node plane strain quadrilateral element with quadratic interpolation and reduced integration (\textsf{CPE8RH}) for both film and substrate layers. We also applied periodic boundary conditions on the left and right edges of the domain. The minimum size of the mesh element is lower than the height of the film. 
We took 12,800 elements, and the numerical results converged and were consistent with the theoretical solutions.


\section*{Appendix C: Coefficients in asymptotic expressions}\refstepcounter{section}\label{asym}


\setcounter{equation}{0}
\setcounter{figure}{0}
\renewcommand{\theequation}{B.\arabic{equation}}
\renewcommand{\thefigure}{B.\arabic{figure}}

The coefficients of the higher-order terms in \cref{eq:nHlam} are given below:

\begin{subequations}
    \begin{equation}
    \begin{split}
    c_4 &= \frac{\pazocal{C}\left(45+\nu_s c_{4a} + \nu_f c_{4b} + \nu_f^2 c_{4c}\right)}{20 \times 2^{1 / 3} \times 3^{2 / 3}(\nu_f - \nu_s)^2(-1+\nu_s)(-3+4 \nu_s)^{4 / 3}}, \qquad
    c_5 = \frac{\pazocal{C}^2(-1+\nu_f)(-1 + 2 \nu_s)(-3 -4 \nu_f + 4 \nu_s)}{2^{2/3} \times 3^{1/3}(\nu_f - \nu_s)^2(-3 + 4 \nu_s)^{5 / 3}}, \\
    c_6 &= \frac{4725 - 2\nu_f^4 c_{6a} -\nu_s c_{6b} + \nu_{f}^3 c_{6c} + \nu_{f}^2 c_{6d} + 2\nu_{f} c_{6e}}{12600(3-4 \nu_s)^2(-1+\nu_s)^2(-\nu_f+\nu_s)^3}, \qquad
    \pazocal{C} = \left((1-\nu_f)(\nu_s-1)\right)^{1/3},
    \end{split}
    \label{eq:nHlam2}
    \end{equation}
and 
 \begin{equation} 
\begin{split}
    c_{4a} &= -64+(13-4 \nu_s) \nu_s, \qquad
    c_{4b} = -71+\nu_s(66+(51-26 \nu_s)\nu_s), \qquad
    c_{4c} = 56-92 \nu_s + 26 \nu_s^2, \\
    c_{6a} &= 33727+2 \nu_s(-73474 +\nu_s(122601 +\nu_s(-92584+26681 \nu_s))), \\
    c_{6b} &= 4410 + \nu_s (82574  + \nu_s(-288226 + \nu_s(399999-260716\nu_s +67034 \nu_s^2))), \\
    c_{6c} &= 95848 + 2\nu_s(-138872 + \nu_s(46253 +\nu_s(235988+\nu_s(-297317+106724 \nu_s)))), \\
    c_{6d} &= -3509 +\nu_s(-202720+\nu_s(841376  +\nu_s(-1218634 +\nu_s(643241+2(22435-53362 \nu_s)\nu_s)))), \\  
    c_{6e} &= -11970+\nu_s(78479+\nu_s(-137927+ \nu_s(-12652 +\nu_s(271583+\nu_s(-276527+89714\nu_s))))).
    \end{split}
        \label{eq:nHlam3}
    \end{equation}
\end{subequations}
Similarly, the coefficients of higher-order terms in the asymptotic expression for critical wavenumber (\cref{eq:nHlam}) are
\begin{subequations}
    \begin{equation}
    \begin{split}
    d_3 &= \frac{4+2 \nu_s-11 \nu_s^2+\nu_f\left(11-32 \nu_s+26 \nu_s^2\right)}{15\left(3-7 \nu_s+4 \nu_s^2\right)}, \qquad
    d_4 = \frac{\left(\frac{2}{3}\right)^{2 / 3}\pazocal{C}(1+2 \nu_f)(-1+2 \nu_s)}{(-3+4 \nu_s)^{4 / 3}}, \\
    d_5 &= \frac{\pazocal{C}^{2} \left(-6761 + \nu_s d_{5a} + 4 \nu_f^2 d_{5b} - 2 \nu_{f} d_{5c} \right)}{3150 \times 2^{2/3} \times 3^{1/3} (-1 + \nu_f) (-1 + \nu_s)^3 (-3 + 4 \nu_s)^{
 5/3}}, 
    \end{split}
    \label{eq:nHkh1}
    \end{equation}
    and 
\begin{equation}
    \begin{split}
     d_{5a} &= 21724 + \nu_s (-19146 + \nu_s (-2636 + 7519 \nu_s)), \quad
 d_{5b} = 646 + \nu_s (-4964 + \nu_s (12906 + \nu_s (-14204 + 5791 \nu_s))), \\
 d_{5c} &= -3821 + \nu_s (7864 + \nu_s (5844 + \nu_s (-22796 + 13609 \nu_s))).
     \end{split}
    \label{eq:nHkh2} 
    \end{equation}
\end{subequations}


\section*{Declaration of Competing Interests}


The authors declare that they have no known competing financial interests or personal relationships that could have appeared to influence the work reported in this paper.


\section*{Acknowledgments}

\noindent
\begin{minipage}{0.1\textwidth}
    \rotatebox{90}{\includegraphics[width=0.85\linewidth]{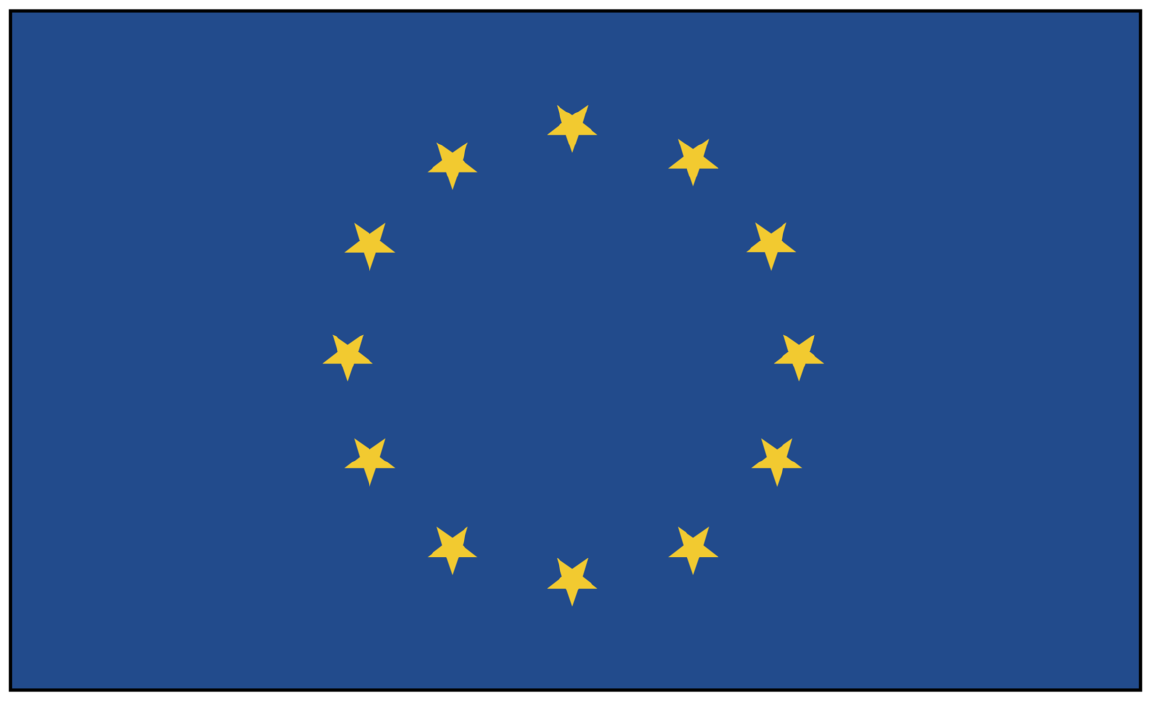}}
\end{minipage}
\hspace{0.5em}
\begin{minipage}{0.85\textwidth}
    This project has received funding from the European Union’s Horizon 2020 research and innovation programme under the Marie Skłodowska-Curie Grant Agreement No. 956401.
\end{minipage}


\section*{Supplementary data}


There is no Supplementary material attached to this article.



\section*{Declaration of Generative AI and AI-Assisted Technologies in the Writing Process}

During the preparation of this work, the authors used the ChatGPT-4o model to correct grammatical mistakes in the original text and to ensure that the paragraph structure was clear and cohesive. After using this tool, the authors reviewed and edited the content as needed and take full responsibility for the content of the published article.




\bibliographystyle{elsarticle-harv-doichange}

\bibliography{ref-list2}

\end{document}